\documentstyle[11pt,epsf,fleqn]{article}
\begin{document}
\title{Electroweak phase transition in an extension of the standard model with a real Higgs singlet}
\author{S.W. Ham$^{(1)}$, Y.S. Jeong$^{(2)}$, and S.K. Oh$^{(1,2)}$
\\
\\
{\it $^{(1)}$ Center for High Energy Physics, Kyungpook National University}\\
{\it Daegu 702-701, Korea} \\
{\it $^{(2)}$ Department of Physics, Konkuk University, Seoul 143-701, Korea}
\\
\\
}
\date{}
\maketitle
\begin{abstract}
The Higgs potential of the standard model with an additional real Higgs singlet is studied
in order to examine if it may allow the strongly first order electroweak phase transition.
It is found that there are parameter values for which this model at the one-loop level
with finite temperature effect may allow the desired phase transition.
Those parameter values also predict that the masses of the neutral scalar Higgs bosons of the model
are consistent with the present experimental bound, and that their productions in $e^+e^-$ collisions
may be searched at the proposed ILC with $\sqrt s$ = 500 GeV in the near future.
\end{abstract}
\vfil
\eject

\section{Introduction}

The possibility of baryogenesis by means of electroweak phase transition has
recently been widely examined, since the electroweak baryogenesis can
in principle be tested in the future accelerator experiments [1].
If the electroweak phase transition is strongly first order,
it can fulfil the departure from thermal equilibrium which is one of
the three conditions required by Sakharov that are necessary
for the dynamic generation of the baryon asymmetry
during the evolution of the universe [2].
It has already been observed that in the standard model (SM) the
electroweak phase transition cannot be strongly first order unless
the mass of the scalar Higgs boson is smaller than its lower bound
set experimentally by LEP [3].
The sufficient strength of the first order electroweak phase transition
is essential for preserving the generated baryon asymmetry at the electroweak
scale. In the literature, a number of articles have been devoted
to study the possibility of accommodating the strongly first order
electroweak phase transition in various models beyond the SM [4].

Among them, an interesting possibility has been investigated
several years ago, where an extension of the SM with a real Higgs
singlet field has been adopted within the context of the
electroweak phase transition [5].
We consider that model is inspiring, because adding a real Higgs singlet field is the
simplest extension of the Higgs sector of the SM.
In that model, the strength of the first order electroweak phase transition has
been stronger than that in the case of the SM.
Due to the presence of cubic terms in the tree-level Higgs potential, a strongly first
order electroweak phase transition has been ensured for a scalar
Higgs boson mass of around 100 GeV, for the top quark mass of 127 GeV.

In this paper, we would like to reexamine the model in a more rigorous way,
in the sense that we perform the full integration of the finite-temperature effective potential
instead of approximating it up to second order in temperature,
that we include the Higgs contributions to the finite temperature Higgs potential
at the one-loop level, and that we take into account the zero-temperature one-loop Higgs potential.
As the experimental lower bound on the scalar Higgs boson mass by now is improved
to be more than 114 GeV, and the top quark mass is usually set as about 175 GeV,
we would like to explore the parameter space of the model, with these new numbers in mind,
where the first order electroweak phase transition may be strong enough.
We find that the model is able to produce strongly first order electroweak phase transition
for some parameter regions where a scalar Higgs boson may have a mass larger than 114 GeV.
Further, we calculate cross sections for scalar Higgs productions in $e^+ e^-$ collisions.

\section{The Higgs Potential}

Let us consider the Higgs potential proposed by Ref. [5] where one real Higgs singlet field
is introduced to the SM Higgs sector.
We follow the notations of Ref. [5].
Thus, the Higgs potential consists of one complex Higgs doublet field $\phi$
and one real Higgs singlet field $S$.
The tree-level Higgs potential is given by
\begin{equation}
    V^0 (\phi, S)
    = \lambda_{\phi} (\phi^{\dagger} \phi)^2
                - \mu_{\phi}^2 \phi^{\dagger} \phi
                + {\lambda_S \over 4} S^4 - {\mu_S^2 \over 2} S^2
                - {\alpha \over 3} S^3 + 2 \lambda (\phi^{\dagger} \phi) S^2
                - {\omega \over 2}(\phi^{\dagger} \phi) S \ ,
\end{equation}
where $\alpha$, $\omega$, $\mu_{\phi}$, and $\mu_S$ are parameters with mass dimension,
and the rest are dimensionless.
We assume that $\lambda_{\phi}, \lambda_S$, and $\lambda$ have a positive values.
The real part of the neutral component of the Higgs doublet field is, as in the SM,
going to develop the non-zero vacuum expectation value (VEV) $u = \sqrt{2} \langle \phi \rangle$
while the Higgs singlet field $v = \langle S \rangle$.
The tree-level values for $u$ and $v$ are determined by the extremum conditions
\begin{equation}
    {\partial V^0 (u,v) \over \partial u} = {\partial V^0 (u,v) \over \partial v} = 0
\end{equation}
where
\begin{equation}
    V^0 (u,v)
    = {\lambda_{\phi} \over 4} u^4 - {\mu_{\phi}^2 \over 2} u^2
             + {\lambda_S \over 4} v^4 - {\mu_S^2 \over 2} v^2
             - {\alpha \over 3} v^3 + {\lambda \over 2} u^2 v^2
             - {\omega \over 4}u^2 v \ .
\end{equation}
The above extremum conditions are satisfied if
\begin{eqnarray}
    \mu_{\phi}^2 & = & \lambda_{\phi} u^2 + \lambda v^2 - {\omega \over 2} v  \ , \cr
    \mu_S^2 & = & \lambda_S v^2 + \lambda u^2 - {\omega \over 4 v} u^2 - \alpha v  \ .
\end{eqnarray}
We denote the zero temperature values of $u$ and $v$ as $u_0$ and $v_0$, respectively.
We take $u_0 = 246$ GeV whereas $v_0$ is not fixed but variable.

In this model, there are two neutral scalar Higgs bosons.
At the tree level, the masses for them are given by
\begin{eqnarray}
    (m^0_{\phi})^2, \ (m^0_S)^2
    & = & {1 \over 2} \left [ - \mu_{\phi}^2 - \mu_S^2 +
        (3 \lambda_{\phi} + \lambda) u^2 + (3 \lambda_S + \lambda) v^2 -
        \left ({\omega \over 2} + 2 \alpha \right) v \right ] \cr
    & &\mbox{}
    \pm \left [ {1 \over 4}
        \left \{ - \mu_{\phi}^2 + \mu_S^2 + (3 \lambda_{\phi} - \lambda) u^2
    - (3 \lambda_S - \lambda) v^2 - \left ({\omega \over 2} - 2 \alpha \right ) v \right \}^2
    \right. \cr
    & &\mbox{}
    \left. + \left (2 \lambda v - {\omega \over 2} \right )^2 u^2 \right ]^{1/2}      \ .
\end{eqnarray}
We associate the lighter neutral scalar Higgs boson with the real part
of the neutral component of the Higgs doublet $\phi$ while the heavier one
with the real Higgs singlet $S$.

At the tree level, the gauge bosons and the top quark mass are $m_W (u_0) = 80.423$ GeV,
$m_Z (u_0) = 91.187$ GeV, and $m_t (u_0) = 175$ GeV. They are all defined in terms of $u_0$,
independently of $v_0$.

Now, we consider the one-loop contributions.
At the one loop level, the zero temperature effective potential [6] and
the finite temperature one [7] are respectively given by
\begin{equation}
    V^1(u,v)
    = \sum_i {n_i  \over 64 \pi^2}  m_i^4(u,v)
           \left[ \log {m_i^2 (u,v) \over \Lambda^2}  - {3 \over 2} \right] \ ,
\end{equation}
\begin{equation}
    V^T(u,v)
    = \sum_i {n_i T^4 \over 2 \pi^2} \int_0^{\infty} dx \ x^2 \
            \log \left [1 \pm \exp {\left ( - \sqrt {x^2+{m_i^2(u,v)/T^2 }} \right )  } \right ] \ ,
\end{equation}
where $n_W = 6$, $n_Z = 3$, $n_t = - 12$, $n_{\phi} = n_S = 1$ account for the degrees
of freedom of each participating particles.
The negative sign in $V^T$ is for bosons while the positive sign for fermions.

In $V^1$, we take the renormalization scale as $\Lambda = 246$ GeV.
The $\Lambda$ dependence of the potential would disappear if all of higher-order contributions
be taken into account.
This is desirable since physical observable quantities should be independent of the renormalization scale.
As we consider only the one-loop contribution in $V^1$, the scale dependence remains.

In order to reduce the effect of the scale dependence, one may introduce appropriate counter terms
to the one-loop contribution, after regularizing the relevant fields in the effective potential.
In our model, the scale dependence on the mass of top quark or on the mass of weak gauge bosons can
completely be eliminated in the one-loop contribution.
This is possible because the Higgs singlet in our model does not contribute to fermion masses
or the weak gauge boson masses.
Thus, the one-loop contributions to these masses in our model are exactly identical to those in the SM.
Hence, the scale dependence of the one-loop contributions to these masses may disappear if
$m^4_i(u) \log(\Lambda^2)$ in Eq. (6) is substituted
by $m^4_i(u)\log(m^2_i(u_0)) -2m^2_i(u) m^2_i(u_0) \approx m^4_i(u) \log(m^2_i(u_0))$,
where we can neglect $2m^2_i(u) m^2_i(u_0)$.
This is effectively equivalent to replacing $\Lambda$ by $m_i(u_0)$.

The case of the Higgs boson mass is more complicated than the case of the fermion masses or
the weak boson masses, since the squared mass of the Higgs boson at the tree level depends
not only $u$ but also $v$, and it is given by Eq. (5) which is not in the form of $C_1+C_2 u^2$
(where $C_1$ and $C_2$ are constants).
Nevertheless, comparing with the one-loop contribution to the top quark mass,
we might effectively set $\Lambda$ in the one-loop contribution to the Higgs boson mass
by either $m^0_{\phi}(u_0, v_0)$ or $m^0_S (u_0, v_0)$,
in a similar way as $\Lambda \rightarrow m_t (u_0)$ in the case of top quark mass.

If a model is related to a new physics at certain scale, the renormalization scale might
be taken comparable to the energy scale of the new physics,
in order to consider the higher-order effects of the model reasonably.
For example, in a supersymmetric axion model, $\Lambda$ is taken to be $10^{11}$ GeV
in the one-loop effective potential, where new physics is expected to emerge at that energy scale [8].
In our study, we are interested in the tree-level mass of the Higgs boson
at the electroweak symmetry breaking scale.
Thus, in our case, we might set $\Lambda = 246$ GeV.

For small $m_i/T$, that is, at high temperature, $V^T$ can be expanded into the well-known form
for the temperature
induced effective potential.
Rather, we perform the integration without using the high temperature approximation in our analysis,
in order to consider relatively large scalar Higgs boson mass.

The full effective potential that we consider therefore consists of
\begin{equation}
    V(u,v) = V^0 (u,v) + V^1(u,v) + V^T (u,v) \ .
\end{equation}
Note that the effective Higgs potential possesses a symmetry under interchange $u \leftrightarrow -u$.
Thus, we may confine ourselves within the region of $u \ge 0$.

In order for the first order electroweak phase transition to take place, we need two separate vacua.
The vacua are defined as the points in the ($u$, $v$)-plane at which the Higgs potential has
local minima.
The two extremum conditions for the Higgs potential may be expressed as
\begin{equation}
{\partial V(u,v) \over \partial u} = f_0 (u,v) = u f_1(u,v) = 0 \ , \quad
{\partial V(u,v) \over \partial v} = f_2 (u,v) = 0 \ .
\end{equation}
where the first equation is decomposed such that it is explicitly satisfied by $u = 0$.

In the above expressions, $f_1 (u, v)$ and $f_2 (u, v)$ are given as
\begin{eqnarray}
f_1 (u,v) & = & - \mu_{\phi}^2 + \lambda_{\phi} u^2 + \lambda v^2 - {\omega v \over 2} + {3 m_W^4(u_0) u^2 \over 8 \pi^2 u_0^2} \left \{\log \left ({m_W^2 (u_0) u^2 \over u_0^2 \Lambda^2} \right ) - 1 \right \}   \cr
& &\mbox{} + {3 m_Z^4(u_0) u^2 \over 16 \pi^2 u_0^2} \left \{\log \left ({m_Z^2 (u_0) u^2 \over u_0^2 \Lambda^2} \right ) - 1 \right \} \cr
& &\mbox{} - {3 m_t^4(u_0) u^2 \over 4 \pi^2 u_0^2} \left \{\log \left ({m_t^2 (u_0) u^2 \over u_0^2 \Lambda^2} \right ) - 1 \right \} \cr
& &\mbox{} + {(3 \lambda_{\phi} + \lambda) \over 32 \pi^2}  \left \{m_{\phi^0}^2 \left (\log  {m_{\phi^0}^2 \over \Lambda^2} - 1 \right ) + m_{S^0}^2 \left (\log {m_{S^0}^2 \over \Lambda^2} - 1 \right ) \right \} \cr
& &\mbox{}  - {\Delta_1 \over 32 \pi^2 u} f (m_{\phi^0}^2, \ m_{S^0}^2) - {3 T^2 m_W^2 (u_0) \over \pi^2 u_0^2} \int_0^{\infty} dx \ x^2 d^-_1(m_W^2)  \cr
& &\mbox{} - {3 T^2 m_Z^2 (u_0) \over 2 \pi^2 u_0^2} \int_0^{\infty} dx \ x^2 d^-_1(m_Z^2) - {6 T^2 m_t^2 (u_0) \over \pi^2 u_0^2} \int_0^{\infty} dx \ x^2 d^+_1(m_t^2) \cr
& &\mbox{} - {T^2 \over 4 \pi^2} \int_0^{\infty} dx \ x^2 d^-_1(m_{\phi^0}^2) \left \{ (3 \lambda_{\phi} + \lambda) - {\Delta_1 \over (m_{S^0}^2 - m_{\phi^0}^2) u} \right \} \cr
& &\mbox{} - {T^2 \over 4 \pi^2} \int_0^{\infty} dx \ x^2 d^-_1(m_{S^0}^2) \left \{ (3 \lambda_{\phi} + \lambda) + {\Delta_1 \over (m_{S^0}^2 - m_{\phi^0}^2) u} \right \} \cr
f_2 (u, v) & = & - \mu_S^2 v + \lambda_S v^3 + \lambda v u^2 - {\omega u^2 \over 4} - \alpha v^2 + {1 \over 32 \pi^2} \left \{(3 \lambda_S + \lambda) v - {\omega \over 4} - \alpha \right \} \cr
& &\mbox{}\times \left \{m_{\phi^0}^2 \left (\log  {m_{\phi^0}^2 \over \Lambda^2} - 1 \right ) + m_{S^0}^2 \left (\log {m_{S^0}^2 \over \Lambda^2} - 1 \right ) \right \} - {\Delta_2 \over 32 \pi^2} f (m_{\phi^0}^2, \ m_{S^0}^2) \cr
& &\mbox{} - {T^2 \over 4 \pi^2} \int_0^{\infty} dx \ x^2 d^-_1(m_{\phi^0}^2) \left \{ (3 \lambda_S + \lambda) v - {\omega \over 4} - \alpha - {\Delta_2 \over (m_{S^0}^2 - m_{\phi^0}^2)} \right \} \cr
& &\mbox{} - {T^2 \over 4 \pi^2} \int_0^{\infty} dx \ x^2 d^-_1(m_{S^0}^2) \left \{ (3 \lambda_S + \lambda) v - {\omega \over 4} - \alpha + {\Delta_2 \over (m_{S^0}^2 - m_{\phi^0}^2)} \right \}     \ ,
\end{eqnarray}
with
\begin{equation}
    f(a, b)
    = {1 \over (b-a)} \left \{a \log  {a \over \Lambda^2}
    - b \log {b \over \Lambda^2} \right \} + 1 \ ,
\end{equation}
\begin{eqnarray}
\Delta_1 & = & \left(2 \lambda_{\phi} u^2 - 2 \lambda_S v^2 - {\omega u^2 \over 4 v} + \alpha v \right) (3 \lambda_{\phi}-\lambda) u + 2 \left(2 \lambda v - {\omega \over 2} \right)^2 u \ , \cr
\Delta_2 & = &\mbox{} - \left(2 \lambda_{\phi} u^2 - 2 \lambda_S v^2 - {\omega u^2 \over 4 v} + \alpha v \right) \left \{(3 \lambda_S-\lambda) v +{\omega \over 4} - \alpha \right \} \cr
& &\mbox{} + 4 \lambda \left(2 \lambda v - {\omega \over 2} \right) u^2 \ , \cr
d_1^{\pm} (m_i^2) & = & \mbox{} - {\exp {\left ( - \sqrt {x^2+{m_i^2/T^2 } } \right ) } \over \sqrt{x^2 + m_i^2/T^2} \left \{ 1 \pm \exp {\left ( - \sqrt {x^2+{m_i^2/T^2 } } \right ) } \right \} }
 \ .
\end{eqnarray}
Note that the negative sign in $d_1^{\pm} (m_i^2)$ is for bosons whereas the positive sign is
for fermions.

Since the first extremum equation is satisfied by either $u = 0$ or $f_1(u,v) = 0$,
one of the two points in the ($u$, $v$)-plane which satisfy the two extremum equations is
simply (0, $v_1$) where $v_1$ is the solution of $f_2(0,v_1) = 0$.
This is the vacuum of the unbroken phase state.
We solve $f_2(0,v_1) = 0$ to obtain $v_1$ numerically by the bisection method.
Let the other point in the ($u$, $v$)-plane which satisfies the two extremum equations be
denoted as ($u_2$, $v_2$), where $u_2 \ne 0$.
This is the vacuum of the broken phase state.
This point is obtained by the Newtons's method.

In order to check that the extremum $(u_2, v_2)$ thus obtained is minimum indeed,
we need a Jacobian matrix
\begin{equation}
    J_{ij} (u, v)
    = \left \lgroup \matrix
    {{\displaystyle {\partial f_1 (u, v) \over \partial u}},
    {\displaystyle {\partial f_1 (u, v) \over \partial v}}  \cr
    {\displaystyle {\partial f_2 (u, v) \over \partial u}},
    {\displaystyle {\partial f_2 (u, v) \over \partial v}}}
    \right \rgroup  \ .
\end{equation}
This point is a minimum of the Higgs potential if they satisfy $J_{ii} (i = 1, 2) > 0$
and $\det (J_{ij}) > 0$.
The distance between the two minima in the ($u$, $v$)-plane is defined
as $v_c = \sqrt{(u_2)^2 + (v_2 -v_1)^2}$.
The nonzero $v_c$ ensures that the electroweak phase transition is discontinuous, that is, first order.
The strength of the first order electroweak phase transition is measured
by the critical temperature $T_c$, which is defined as the temperature at
which the values of the potential at the two minima are equal: $V(0,v_1) = V(u_2,v_2)$.
We determine $T_c$ by varying temperature $T$ until $V(0,v_1) = V(u_2,v_2)$ is met.
If the ratio  $v_c/T_c$ is larger than 1, the first order electroweak phase transition
is defined conventionally as a strong one.

Now, let us evaluate the masses of the two neutral scalar Higgs bosons from the symmetric
$2 \times 2$ mass matrix at the one-loop level.
The masses are evaluated at zero temperature.
Thus, $V^T$ does not contribute to the masses at the one-loop level, and we should take
$u = u_0$ and $v = v_0$ in the final expressions.
The elements of the mass matrix for the neutral scalar Higgs bosons are given by
\begin{eqnarray}
    M_{11} & = & (M_{11}^0 + M_{11}^1)   \ , \cr
    M_{22} & = & (M_{22}^0 + M_{22}^1)   \ , \cr
    M_{12} & = & (M_{12}^0 + M_{12}^1)   \ ,
\end{eqnarray}
where $M_{ij}^0$ and $M_{ij}^1$ are obtained from $V^0$ and $V^1$, respectively.
Explicitly, they are obtained as
\begin{eqnarray}
    M_{11}^0 & = & 2 \lambda_{\phi} u^2  \ , \cr
    M_{22}^0 & = & 2 \lambda_S v^2 - \alpha v + {\omega u^2 \over 4 v} \ , \cr
    M_{12}^0 & = & 2 \lambda u v - {\omega \over 2} u \ .
\end{eqnarray}
and
\begin{eqnarray}
M_{11}^1 & = & {3 m_W^4 (u) \over 4 \pi^2 u^2} \log \left ({m_W^2 (u) \over \Lambda^2 } \right )  + {3 m_Z^4 (u) \over 8 \pi^2 u^2} \log \left ({m_Z^2 (u) \over \Lambda^2 } \right )  - {3 m_t^4 (u) \over 2 \pi^2 u^2} \log \left ({m_t^2 (u) \over \Lambda^2 } \right ) \cr
& &\mbox{} + {\Delta_1^2 \over 32 \pi^2} {g (m_{\phi^0}^2, \ m_{S^0}^2) \over (m_{S^0}^2 - m_{\phi^0}^2)^2} - {(3 \lambda_{\phi} - \lambda)^2 u^2 \over 16 \pi^2 v^2} f (m_{\phi^0}^2, \ m_{S^0}^2) \cr
& &\mbox{} + {(3 \lambda_{\phi} + \lambda)^2 u^2 \over 32 \pi^2} \log \left ({m_{\phi^0}^2  m_{S^0}^2  \over \Lambda^4} \right )
+ {(3 \lambda_{\phi} + \lambda) u \Delta_1 \over 16 \pi^2} {\log (m_{S^0}^2 / m_{\phi^0}^2) \over (m_{S^0}^2 - m_{\phi^0}^2)} \ , \cr
M_{22}^1 & = & {1 \over 32 \pi^2} \left ({\omega \over 4 v} + {\alpha \over v} \right ) \left \{m_{\phi^0}^2 \left (\log  {m_{\phi^0}^2 \over \Lambda^2} - 1 \right ) + m_{S^0}^2 \left (\log {m_{S^0}^2 \over \Lambda^2} - 1 \right ) \right \} \cr
& &\mbox{} + {\Delta_2^2 \over 32 \pi^2} {g (m_{\phi^0}^2, \ m_{S^0}^2) \over (m_{S^0}^2 - m_{\phi^0}^2)^2} - {\Delta_3 \over 32 \pi^2} f (m_{\phi^0}^2, \ m_{S^0}^2) \cr
& &\mbox{} + {1 \over 32 \pi^2} \left \{(3 \lambda_S + \lambda) v - {\omega \over 4} - \alpha \right\}^2 \log \left ({m_{\phi^0}^2 m_{S^0}^2 \over \Lambda^4} \right )  \cr
& &\mbox{} + {\Delta_2 \over 16 \pi^2} \left \{(3 \lambda_S + \lambda) v - {\omega \over 4} - \alpha \right\} {\log (m_{S^0}^2 /  m_{\phi^0}^2) \over (m_{S^0}^2  - m_{\phi^0}^2)} \ , \cr
M_{12}^1 & = & {\Delta_1 \Delta_2 \over 32 \pi^2} {g (m_{\phi^0}^2, \ m_{S^0}^2) \over (m_{S^0}^2 - m_{\phi^0}^2)^2} - {\Delta_4 \over 32 \pi^2} f (m_{\phi^0}^2, \ m_{S^0}^2) \cr
& &\mbox{} + {(3 \lambda_{\phi} + \lambda) u \over 32 \pi^2} \left \{ (3 \lambda_S + \lambda) v - {\omega \over 4} - \alpha \right \} \log \left ({m_{\phi^0}^2 m_{S^0}^2  \over \Lambda^4} \right ) \cr
& &\mbox{} + {1 \over 32 \pi^2} \left [\left \{ (3 \lambda_S + \lambda) v - {\omega \over 4} - \alpha\right \}\Delta_1 + (3 \lambda_{\phi} + \lambda) u \Delta_2 \right ] {\log (m_{S^0}^2 / m_{\phi^0}^2) \over (m_{S^0}^2 - m_{\phi^0}^2)}  \ ,
\end{eqnarray}
where
\begin{equation}
    g(a, b) = {a + b \over a - b} \log {b \over a} + 2 \ ,
\end{equation}
\begin{eqnarray}
\Delta_3 & = & 2 \left \{ (3 \lambda_S - \lambda) v + {\omega \over 4} - \alpha \right \}^2 + \left (2 \lambda_{\phi} u^2 - 2 \lambda_S v^2 - {\omega u^2 \over 4 v} + \alpha v \right ) \left ({\omega \over 4 v} - {\alpha \over v} \right ) \cr
& &\mbox{} + 2 \lambda \omega {u^2 \over v} \ , \cr
\Delta_4 & = &\mbox{} - 2 (3 \lambda_{\phi} - \lambda) \left \{(3 \lambda_S - \lambda) v + {\omega \over 4} - \alpha \right \} u + 8 \lambda \left (2 \lambda v - {\omega \over 2} \right ) u \ .
\end{eqnarray}

In terms of these matrix elements, the neutral scalar Higgs boson masses at the one-loop level are
given as
\begin{equation}
    m_{\phi}^2, \ m_{S}^2
    = {1 \over 2} \left [(M_{11} + M_{22}) \mp \sqrt{(M_{22}-M_{11})^2 + (M_{12})^2 } \right ] \ ,
\end{equation}
which depend on the quartic coupling coefficients $\lambda_{\phi}$, $\lambda_S$, and $\lambda$,
and three parameters $v_0$, $\omega$, and $\alpha$, having mass dimension.
Here, it is assumed that $m_{\phi} < m_S$.
The quartic coupling coefficients satisfy the condition of
$\lambda_i/4\pi < 1$ ($\lambda_i = \lambda_{\phi}, \lambda_S, \lambda$)
for perturbation theory to work, in principle.
Thus, they are not large.
In practice, we may set their values to be order of 1 or less.

The value of $v_0$ should be comparable to that of $u_0 = 246$ GeV,
because it has a similar property as $u_0$.
Note that $u$ is the Higgs field in the SM for $v =0$ and $u_0$ in the SM is
the electroweak symmetry breaking scale.
Likewise, we assume that the other parameters $\omega$ and $\alpha$ are not so much different from $u_0$.
Since we are interested in the mass region of 200 to 400 GeV
for the upper bound on the Higgs boson masses, we choose the values of the parameters
such that they yield the Higgs boson masses within this region.

For numerical study, we randomly search the parameter space by varying the relevant parameters
within the ranges of $0 < \lambda_{\phi}, \lambda_S, \lambda \le 0.7$, $0 < v_0 \le 300$ GeV,
and $0 < - \omega, \alpha \le 100$ GeV.
Starting with a set of relevant parameters whose values are chosen randomly within the allowed ranges,
we examine if $v_c/T_c > 1$, which is the criterion for the first order electroweak phase transition
to be strong.
If this criterion is met, we continue to calculate the masses of the neutral scalar Higgs bosons.
This sequence of calculations produces a point in the ($m_{\phi}$, $v_c/T_c$)-plane or
in the ($m_S$, $v_c/T_c$)-plane.
In this way, we perform calculations to obtain 5000 points.
Figs. 1a and 1b show the results of our numerical studies on the masses of the neutral scalar Higgs bosons.
In Fig. 1a, the 5000 points thus obtained are plotted in the ($m_{\phi}$, $v_c/T_c$)-plane,
and in Fig. 1b in the ($m_S$, $v_c/T_c$)-plane.
Notice that the points in the ($m_S$, $v_c/T_c$)-plane are scattered slightly toward
to the right of the plane, as compared to the points scattered in the ($m_{\phi}$, $v_c/T_c$)-plane,
since we assume that $m_{\phi} < m_S$.
These figures, therefore, indicate that there are sufficiently wide regions in the parameter space
where the first order electroweak phase transition is strong as well as
the masses of the neutral scalar Higgs bosons are relatively large.

We now examine the possibility of detecting these neutral scalar Higgs bosons
in the future linear $e^+e^-$ collider.
The three main production processes for the neutral scalar Higgs bosons in $e^+ e^-$ collisions are:
\begin{eqnarray}
    &&\mbox{Higgs-strahlung : } e^+e^- \rightarrow Z \phi, \ ZS  \ , \cr
    &&\mbox{$WW$ fusion : } e^+e^- \rightarrow {\bar \nu}_e \nu_e \phi, \  {\bar \nu}_e \nu_e S \ , \cr
    &&\mbox{$ZZ$ fusion : } e^+e^- \rightarrow e^+ e^- \phi, \ e^+ e^- S \ , \nonumber
\end{eqnarray}
where the Higgs-strahlung process is dominant at the center of mass energy corresponding
to that of LEP2 whereas $WW$ and $ZZ$ fusion processes are comparable
at much higher center of mass energy, such as that of the future linear $e^+e^-$ collider
with $\sqrt{s} \ge 500$ GeV.

The cross section for $\phi$ or $S$ production via the Higgs-strahlung process is related
to the corresponding production cross section of the SM scalar Higgs boson, $H$, as
\begin{eqnarray}
    & &\sigma (Z \phi) = \cos^2 (\gamma) \sigma_{\rm SM} (Z H) \ ,  \cr
    & &\sigma (Z S) = \sin^2 (\gamma) \sigma_{\rm SM} (Z H) \ ,
\end{eqnarray}
and, similarly, the cross sections for $\phi$ and $S$ production via the $WW$ fusion process or
the $ZZ$ fusion process are related to the corresponding production cross section
of the SM scalar Higgs boson, $H$, as
\begin{eqnarray}
    &&\sigma (VV \phi) = \cos^2 \gamma \; \sigma_{\rm SM} (VV H)  \ ,  \cr
    &&\sigma (VV S) = \sin^2 \gamma \; \sigma_{\rm SM} (VV H)  \ ,
\end{eqnarray}
where $VV = WW$ or $ZZ$, and $\gamma$ is the mixing angle between the two neutral scalar Higgs bosons:
\begin{equation}
    \cos 2 \gamma = {M_{11} - M_{22} \over \sqrt{(M_{22} - M_{11})^2 + 4 (M_{12})^2}} \ .
\end{equation}
These relationships hold because the coefficients for $VV\phi$ and $VVS$  coupling in our model are
respectively $\cos\gamma$ and $\sin\gamma$ times the $VVH$ coupling coefficient in the SM.

For a given set of parameter values in our model, the above relationships indicate that,
in case of the Higgs-strahlung process, although neither $\sigma(Z\phi)$ nor $\sigma(ZS)$ is
larger than $\sigma_{\rm SM} (ZH)$, they cannot be simultaneously smaller than the SM value.
If one becomes negligible, the other approaches to the SM value.
The same argument holds in case of the fusion processes.
If $\cos\gamma = 1$, the neutral scalar boson $\phi$ in our model would behave like $H$ in the SM.
Its properties would be identical to those for $H$ in the SM.
In particular, if $\cos\gamma =1$, the lower bound on the mass of $\phi$ would be
the same as the lower bound on the SM Higgs boson mass, 115 GeV,
which has been set by the LEP experiments.
If, on the other hand, $\cos\gamma = 0$, the neutral Higgs boson $S$ in our model would behave
like $H$ in the SM.

Let us consider the Higgs-strahlung process.
At the center of mass energy of LEP2, the Higgs-strahlung process in the SM is dominant
over the fusion processes.
We first calculate both $\sigma(Z\phi)$ and $\sigma(ZS)$, and then choose the larger one
\[
    \sigma_{\rm max}^H = {\rm max} \{\sigma (Z \phi), \sigma (Z S)\} \ ,
\]
since at least one of them is comparable to $\sigma_{\rm SM} (ZH)$.
We calculate $\sigma_{\rm max}^H$ for the each of the 5000 sets of relevant parameter values
which produce the points in Figs. 1.
We plot the result against $m_{\phi}$ in Fig. 2.
The number of points in Fig. 2 is smaller than those in Figs. 1, because those points
in Figs. 1 yielding  $m_{\phi} > 118$ GeV are kinematically excluded from Fig. 2.

From Fig. 2, the lower bound on $m_{\phi}$ might be roughly estimated.
It can be seen from Fig. 2 that the lowest value for $m_{\phi}$ is about 20 GeV.
However, this lower bound on $m_{\phi}$ does not take into account the discovery limit.
If the discovery limit of the neutral scalar Higgs boson $\phi$ in our model
via the Higgs-strahlung process in $e^+e^-$ collisions at the LEP2 energy is, for example, 400 fb,
those points above 400 fb should be excluded since LEP2 had not discovered $\phi$ or $S$.
Then, the lower bound on $m_{\phi}$ is pushed upward to about 65 GeV.
It is known that the cross section for the SM Higgs production via the Higgs-strahlung process
in $e^+e^-$ collisions at the LEP2 energy is about 200 to 250 fb for $m_H \approx 110$ GeV.
Assuming that the discovery limit of the $\phi$ production cross section in our model is
roughly the same order of magnitude, namely, a few hundred fb,
we might estimate that $m_{\phi} \ge 60$ GeV.
Similar argument would apply to the lower bound on the mass of the neutral scalar boson $S$,
if we plot  $\sigma_{\rm max}^H$ against $m_S$.

The LEP2 experiments have provided detailed, mass dependent, limits on the Higgs production
and decay cross sections, yielding a limit on the Higgs mass as a function of the ratio of the cross
section times the branching ratio of its decay into $b \bar{b}$ pairs to the SM value.
Combined with similar analyses for other decay modes, the LEP2 experiments may exclude a Higgs boson with
mass of order 60 GeV even if its production cross section is a tenth of the SM value.
In Ref. [9], the same analysis has been done independently of the specific hadronic decay mode of the
Higgs boson.
Comparing the result of Ref. [9] with our values, one can conclude that the bound on $m_{\phi}$ is indeed
of order 60 GeV. Among the points in Fig. 2, those points with $m_{\phi} \ge 60$ GeV
and small cross section, lower than 100 fb, are consistent with LEP data.
However, the key point here is that the LEP2 experiments cannot exclude the whole points in Fig. 2.
This implies that there are parameter regions in our model, where both the electroweak phase transition
is strongly first order and the productions of the neutral scalar Higgs bosons via the
Higgs-strahlung process in $e^+e^-$ collisions are consistent with the LEP2 experiments.

Now, we increase $\sqrt{s}$ of the $e^+e^-$ collisions up to the proposed ILC center
of mass energy of 500 GeV.
At this energy, depending on the scalar Higgs boson mass, the $WW$ and $ZZ$ fusion processes are
comparable to the Higgs-strahlung process.
At this energy, the search for the SM Higgs boson via the Higgs-strahlung process in $e^+e^-$ collision
may extend to the mass region up to about 400 GeV.
The discovery limit of the cross section for the SM Higgs production via the Higgs-strahlung process
at the ILC of $\sqrt{s} = 500$ GeV is about 3 fb.
For our model, taking the mixing angle factor $\cos^2\gamma$ or $\sin^2\gamma$ into account,
the discovery limit would slightly be larger than the SM one.
It is thus possible to assume that the discovery limit of the cross section for the Higgs production
via the Higgs-strahlung process in our model is less than 10 fb.

At $\sqrt{s} = 500$ GeV, all of the 5000 sets of parameter values we select in Figs. 1 are
kinematically allowed to produce $\phi$ or $S$ in $e^+e^-$ collisions.
We calculate the cross sections for $\phi$ and $S$ production via the Higgs-strahlung process,
the $WW$ process, and the $ZZ$ process, namely, $\sigma(Z\phi)$, $\sigma(ZS)$, $\sigma(WW\phi)$,
$\sigma(WWS)$, $\sigma(ZZ\phi)$, and $\sigma(ZZS)$.
We also choose in each channel the larger one between the $\phi$ production cross section
and the $S$ production cross section:
$\sigma_{\rm max}^H = {\rm max} \{\sigma (Z \phi), \sigma (Z S)\}$,
$\sigma_{\rm max}^W = {\rm max} \{\sigma (WW \phi), \sigma (WW S)\}$,
and $\sigma_{\rm max}^Z = {\rm max} \{\sigma (ZZ \phi), \sigma (ZZ S)\}$.

Our results are plotted in Figs. 3-5.
Figs. 3 show the results for the Higgs-strahlung process, Fig. 3a for $\sigma(Z\phi)$,
Fig. 3b for $\sigma(ZS)$, and Fig. 3c for $\sigma_{\rm max}^H$.
For simplicity, all of these figures are plotted against $m_{\phi}$.
Fig. 3c suggests that, at $\sqrt{s} = 500$ GeV, the production cross sections
via the Higgs-strahlung process are perceivably higher than the discovery limit of 10 fb,
for 5000 sets of parameter values in our model.
Therefore, these sets of parameter values in our model may be explored
at the ILC with $\sqrt{s} = 500$ GeV.

Figs. 4 show the results for the $WW$ fusion process, and Figs. 5 for the $ZZ$ fusion process.
The production cross sections via the $ZZ$ fusion process are generally smaller than those
via the other channels.
The production cross sections via the $WW$ fusion process are considerably large for smaller scalar mass,
but rapidly decrease as the scalar mass increases.
These behaviors are expectably consistent with the SM predictions.

\section{Discussions}

Successful baryogenesis requires the presence of CP violating sources beyond the CKM phase
in the SM.
This is because the CKM phase in the SM cannot produce sufficient CP violation
for the required amount of the electroweak baryogenesis.
The Higgs sector of our model cannot accommodate CP violation since
it consists of the SM Higgs doublet and a real Higgs singlet.
Thus, we need sources of CP violation other than its Higgs sector.
As possible sources of CP violation, scenarios for new physics are good candidate.
Assuming the existence of new physics have usually been induced to solve the hierarchy problem.

Among various scenarios for new physics, two of them can be incorporated with our model [10, 11].
In one scenario [10], higher-dimensional interaction terms associated with the Higgs doublet may
be introduced at the electroweak energy scale.
In this case, the CP violation occurs at the electroweak scale, and the effective Yukawa coupling
for top quark is given by $h^{eff}_t = h_t [1+ c_t  e^{i\xi} (\phi^2 -v^2/2) /\Lambda^2_c ]$,
where $\xi$ is a CP violating phase, $c_t$ is a real parameter, and $\Lambda_c$ is
a cut off energy scale related to the new physics which is much larger than the electroweak energy scale.
Ref. [10] suggests that the energy scale for the new physics might be the Majorana mass scale
of the right-handed neutrino in left-right symmetric models.

Another scenario is more realistic and suitable for our model [11], where energy scale
for the electroweak phase transition and the energy scale for the CP violation can be made different.
The Higgs sector of the model studied in Ref. [11] has a conventional complex Higgs doublet
and an extra complex Higgs singlet.
The conventional complex Higgs doublet play the role of the SM Higgs doublet,
introducing the SM Higgs boson, and the extra complex Higgs singlet introduces
two additional neutral Higgs bosons.
Both of the additional neutral Higgs bosons are heavier than the SM Higgs boson; one of them stays
at the electroweak scale whereas the other has a mass of high energy scale.
Thus, among the three neutral Higgs bosons of the model, there are two neutral Higgs bosons
at the electroweak scale and the third one is decoupled from them.

The model considered in Ref. [11] introduces not only the extra complex Higgs singlet
but also extra vector-like down quark.
The introduction of a vector-like quarks to the SM is attractive since they naturally arise
in grand unified theories such as $E_6$, noted Ref. [11].
In the fermion sector of the model considered in Ref. [11], the SM fermions stay at the electroweak scale
whereas the vector-like down quark stays at high energy scale.
Consequently, the electroweak phase transition occurs at the electroweak scale,
whereas the CP violation occurs at high energy scale.
In this case, therefore, the contributions of the heavy third neutral Higgs boson
and the vector-like down quark in the finite-temperature effective potential are Boltzmann suppressed
and thus negligible when the electroweak phase transition are considered at the electroweak scale.
In other words, the model considered in Ref. [11] becomes effectively identical to our model
at the electroweak scale.
Actually, by suitably redefining the tree-level Higgs potential,
and decoupling the third neutral Higgs boson, in the model considered in Ref. [11],
it becomes exactly the tree-level Higgs potential of our model.
Thus, our model for the electroweak phase transition may be extended naturally into the model considered
in Ref. [10] in order to accommodate the CP violation at high energy scale,
without modifying our study at the electroweak scale.

The terms proportional to $\alpha$ and $\omega$ in the tree-level Higgs potential come
from $Z_2$ symmetry breaking terms.
If $\alpha = \omega = 0$, these two terms are absent, and the parameter region allowing strongly
first order electroweak phase transition disappears.
This can be seen, for example, in Fig. 2 of Ref. [11], where $\alpha$ is the same and $\omega$
is $ -4 \xi$.
Thus, the $Z_2$ breaking terms with $\alpha \ne \omega \ne 0$ are crucial
for the strongly first order electroweak phase transition.

Actually, there should be a potential barrier between two vacua in the tree-level Higgs potential
in order for the strongly first order electroweak phase transition to occur.
As an example, in a nonminimal supersymmetric model, a trilinear term may be introduced
in its tree-level Higgs potential to replace the $\mu$-term.
This term allows strongly first order electroweak phase transition
in the nonminimal supersymmetric model.
Also, in Ref. [5], one can see that there are a global minimum and a local minimum
in the tree-level Higgs potential, which become two separate global minima by the contribution
of the temperature dependent effective potential.

\section{Conclusions}

The SM with an additional real Higgs singlet in its Higgs sector is studied
within the context of the electroweak phase transition.
Radiative corrections by the one-loop contribution from gauge bosons, top quark,
and the Higgs bosons are included, as well as the finite temperature effects.
The one-loop finite-temperature corrections are integrated numerically,
instead of employing the high temperature approximation.
By randomly choosing a large number of sets of the relevant parameter values
within the ranges of $0 < \lambda_{\phi}, \lambda_S, \lambda \le 0.7$,
$0 < v_0 \le 300$ GeV, and $0 < - \omega, \alpha \le 100$ GeV,
we select those sets of parameter values that satisfy $v_c/T_c > 1$.
For those particular sets of parameter values,
we calculate the masses of the two neutral scalar Higgs bosons of the model and
the production cross sections for them via three dominant processes in $e^+e^-$ collisions.
We find that there are as much as 5000 sets of parameter values of this model
that may accommodate the strongly first order electroweak phase transition,
which yield reasonably large masses for the neutral scalar Higgs bosons and
large production cross sections in $e^+e^-$ collisions with $\sqrt{s}$ = 500 GeV.
Thus, for certain parameter values, the prediction on the masses of the neutral scalar Higgs bosons
are consistent with the present experimental lower bound.
The results of our calculations suggest that this model may be tested
in the near future at the proposed ILC.

\vskip 0.3 in

\noindent
{\large {\bf Acknowledgments}}
\vskip 0.2 in
\noindent
This work was supported by Korea Research Foundation Grant (2001-050-D00005).

\vskip 0.2 in


\vfil\eject

{\bf Figure Captions}

\vskip 0.3 in
\noindent
Fig. 1a : Within the ranges of $0 < \lambda_{\phi}, \lambda_S, \lambda \le 0.7$, $0 < v_0 \le 300$ GeV,
and $0 < - \omega, \alpha \le 100$ GeV, 5000 sets of parameter values are randomly selected
which yield $v_c/T_c >1$, and they are plotted against $m_{\phi}$.

\vskip 0.3 in
\noindent
Fig. 1b : For the same 5000 sets of parameter values as in Fig. 1a, $v_c/T_c$
against $m_S$ are plotted.

\vskip 0.3 in
\noindent
Fig. 2 : For the same 5000 sets of parameter values as in Fig. 1a, the larger one
of the production cross sections, $\sigma_{\rm max}^H = {\rm max} \{\sigma (Z \phi),
\sigma(Z S)\}$, for the neutral scalar Higgs bosons via Higgs-strahlung process
in $e^+e^-$ collisions with $\sqrt{s} = 209$ GeV is plotted against $m_{\phi}$.

\vskip 0.3 in
\noindent
Fig. 3a : For the same 5000 sets of parameter values as in Fig. 1a, the cross section
for $\phi$ production via Higgs-strahlung process in $e^+e^-$ collisions with $\sqrt{s} = 500$ GeV
is plotted against $m_{\phi}$.

\vskip 0.3 in
\noindent
Fig. 3b : For the same 5000 sets of parameter values as in Fig. 1a, the cross section
for $S$ production via Higgs-strahlung process in $e^+e^-$ collisions with $\sqrt{s} = 500$ GeV
is plotted against $m_{\phi}$.

\vskip 0.3 in
\noindent
Fig. 3c : For the same 5000 sets of parameter values as in Fig. 1a,
the larger one between the cross section for $\phi$ production and the cross section
for $S$ production via Higgs-strahlung process in $e^+e^-$ collisions with $\sqrt{s} = 500$ GeV
is plotted against $m_{\phi}$.

\vskip 0.3 in
\noindent
Figs. 4a-c : The same as Figs. 3a-c, except for $WW$ fusion process instead
of the Higgs-strahlung process.

\vskip 0.3 in
\noindent
Figs. 5a-c : The same as Figs. 3a-c, except for $ZZ$ fusion process instead
of the Higgs-strahlung process.

\vfil\eject

\setcounter{figure}{0}
\def\figurename{}{}%
\renewcommand\thefigure{Fig. 1a}
\begin{figure}[t]
\epsfxsize=12cm
\hspace*{2.cm}
\epsffile{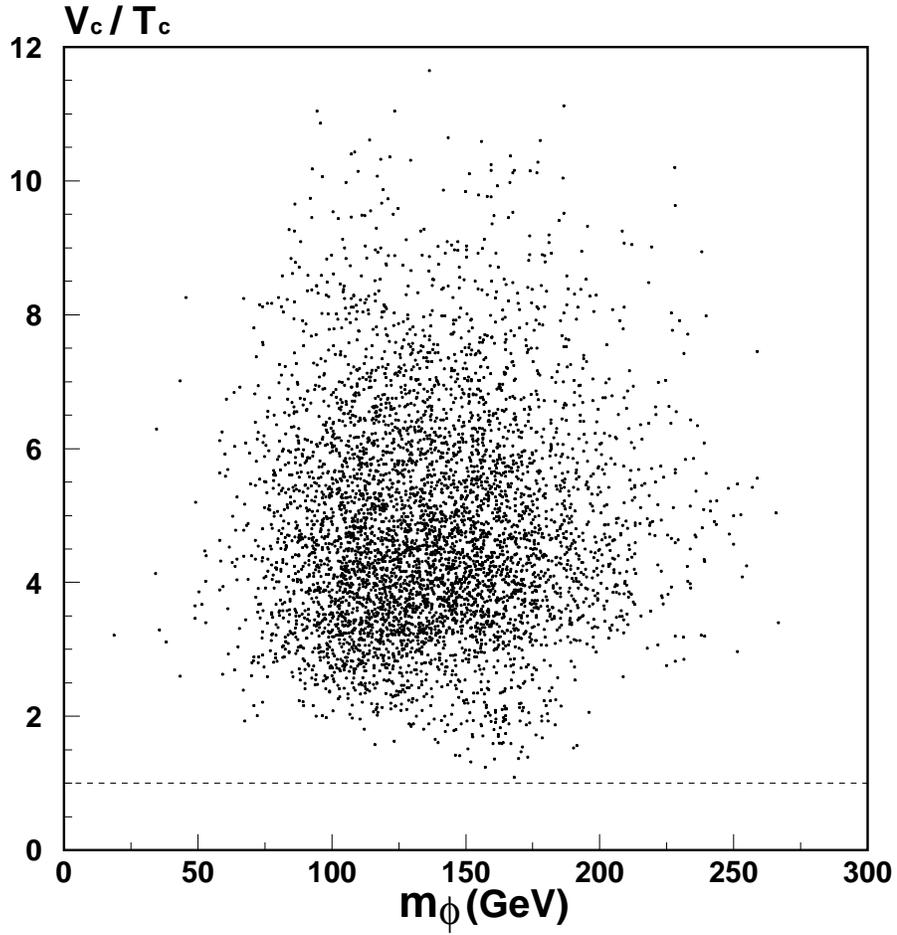}
\caption[plot]{Within the ranges of $0 < \lambda_{\phi}, \lambda_S, \lambda \le 0.7$,
$0 < v_0 \le 300$ GeV, and $0 < - \omega, \alpha \le 100$ GeV,
5000 sets of parameter values are randomly selected which yield $v_c/T_c >1$,
and they are plotted against $m_{\phi}$.}
\end{figure}
\renewcommand\thefigure{Fig. 1b}
\begin{figure}[t]
\epsfxsize=12cm
\hspace*{2.cm}
\epsffile{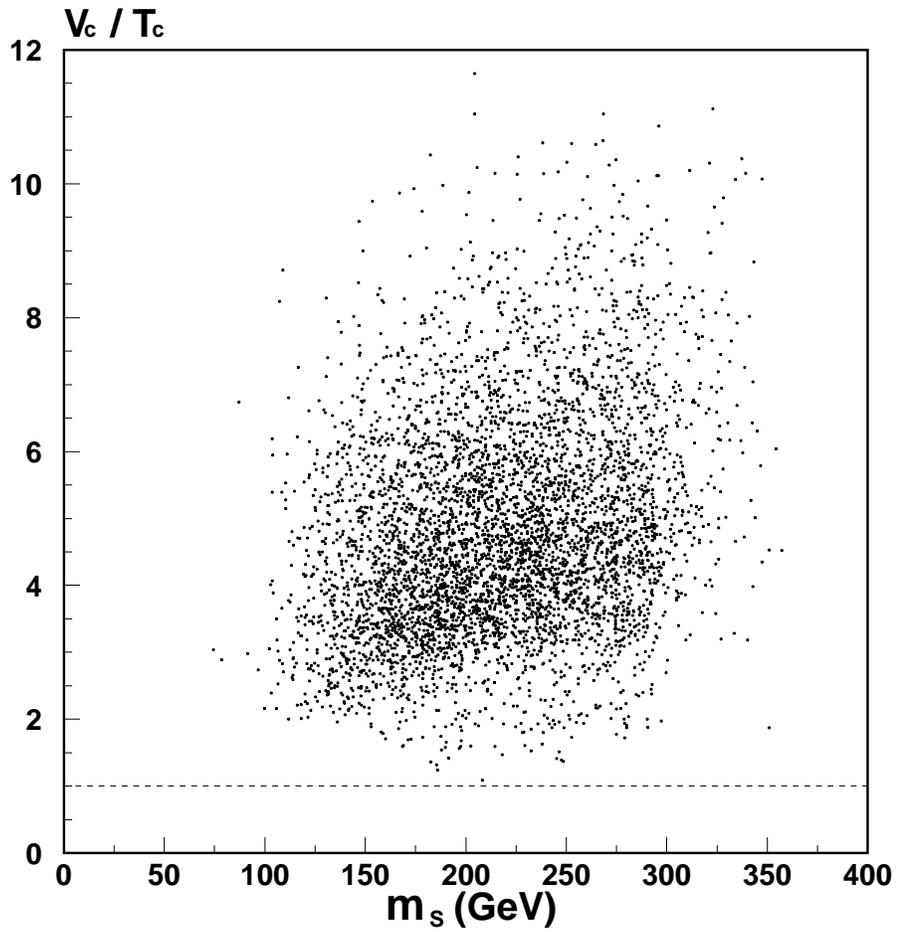}
\caption[plot]{For the same 5000 sets of parameter values as in Fig. 1a, $v_c/T_c$
against $m_S$ are plotted.}
\end{figure}
\renewcommand\thefigure{Fig. 2}
\begin{figure}[t]
\epsfxsize=12cm
\hspace*{2.cm}
\epsffile{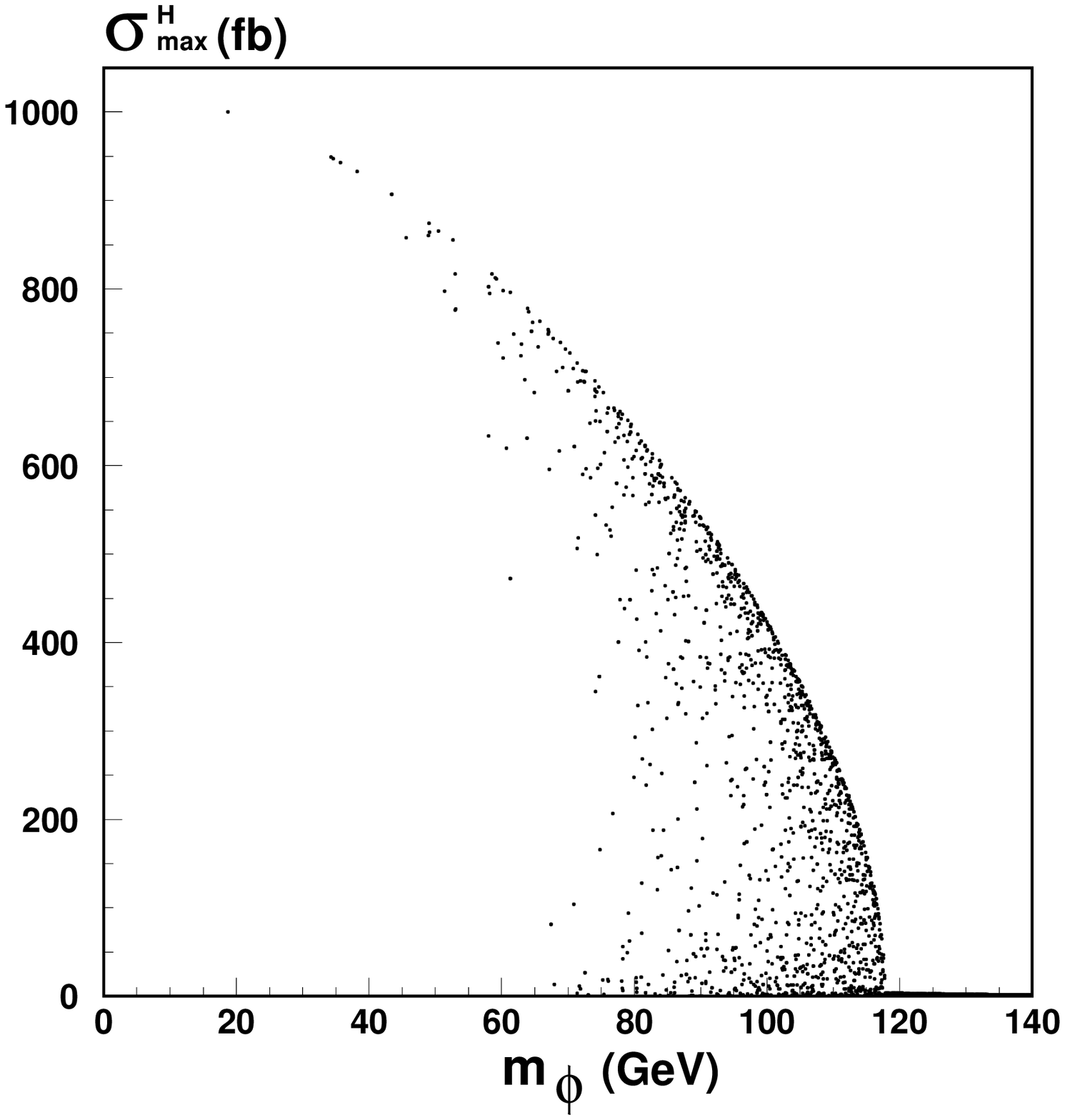}
\caption[plot]{For the same 5000 sets of parameter values as in Fig. 1a,
the larger one of the production cross sections,
$\sigma_{\rm max}^H = {\rm max} \{\sigma (Z \phi), \sigma(Z S)\}$,
for the neutral scalar Higgs bosons via Higgs-strahlung process in $e^+e^-$ collisions
with $\sqrt{s} = 209$ GeV is plotted against $m_{\phi}$.}
\end{figure}
\renewcommand\thefigure{Fig. 3a}
\begin{figure}[t]
\epsfxsize=12cm
\hspace*{2.cm}
\epsffile{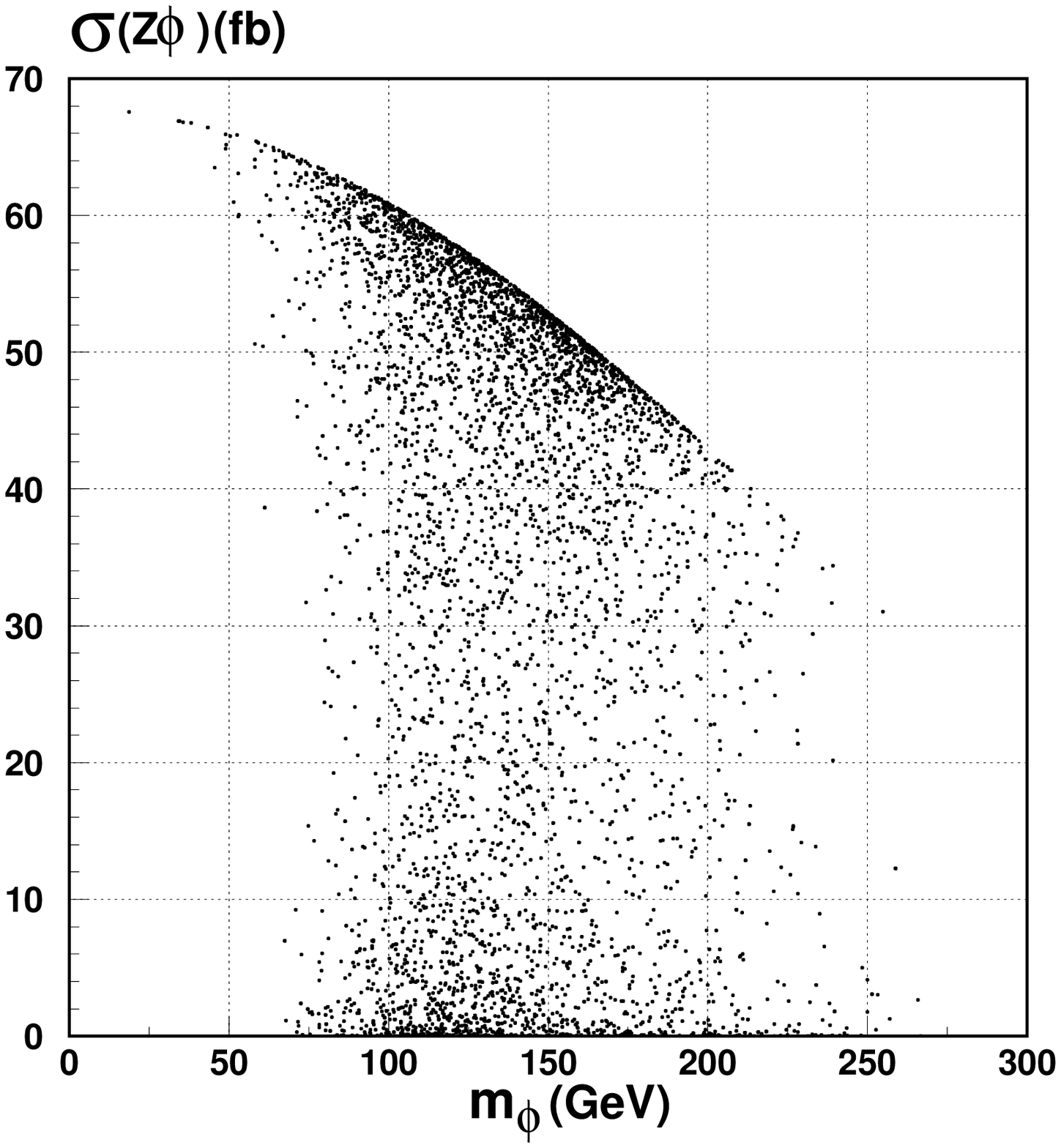}
\caption[plot]{For the same 5000 sets of parameter values as in Fig. 1a,
the cross section for $\phi$ production via Higgs-strahlung process in $e^+e^-$ collisions
with $\sqrt{s} = 500$ GeV is plotted against $m_{\phi}$.}
\end{figure}
\renewcommand\thefigure{Fig. 3b}
\begin{figure}[t]
\epsfxsize=12cm
\hspace*{2.cm}
\epsffile{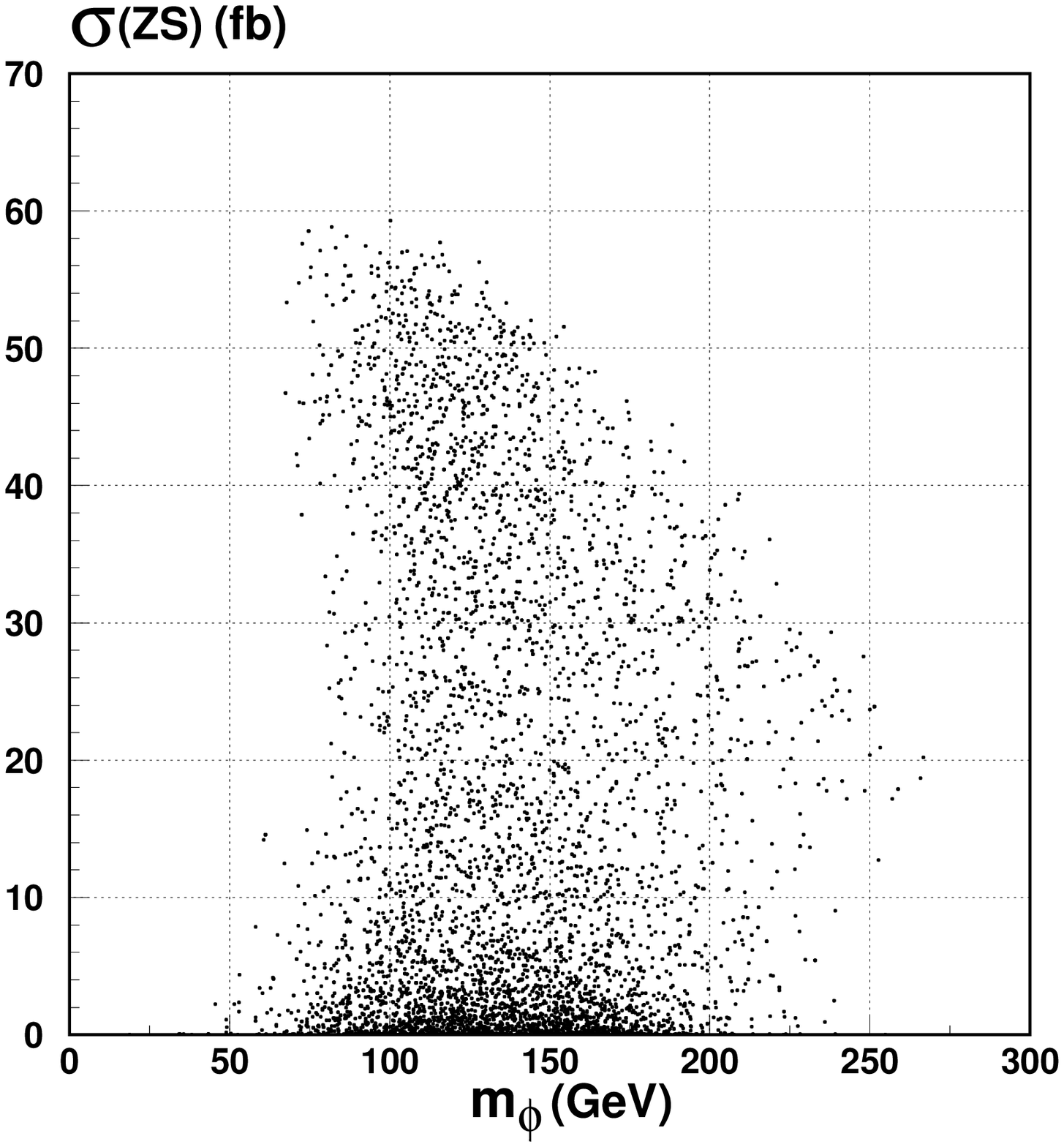}
\caption[plot]{For the same 5000 sets of parameter values as in Fig. 1a, the cross section
for $S$ production via Higgs-strahlung process in $e^+e^-$ collisions with $\sqrt{s} = 500$ GeV
is plotted against $m_{\phi}$.}
\end{figure}
\renewcommand\thefigure{Fig. 3c}
\begin{figure}[t]
\epsfxsize=12cm
\hspace*{2.cm}
\epsffile{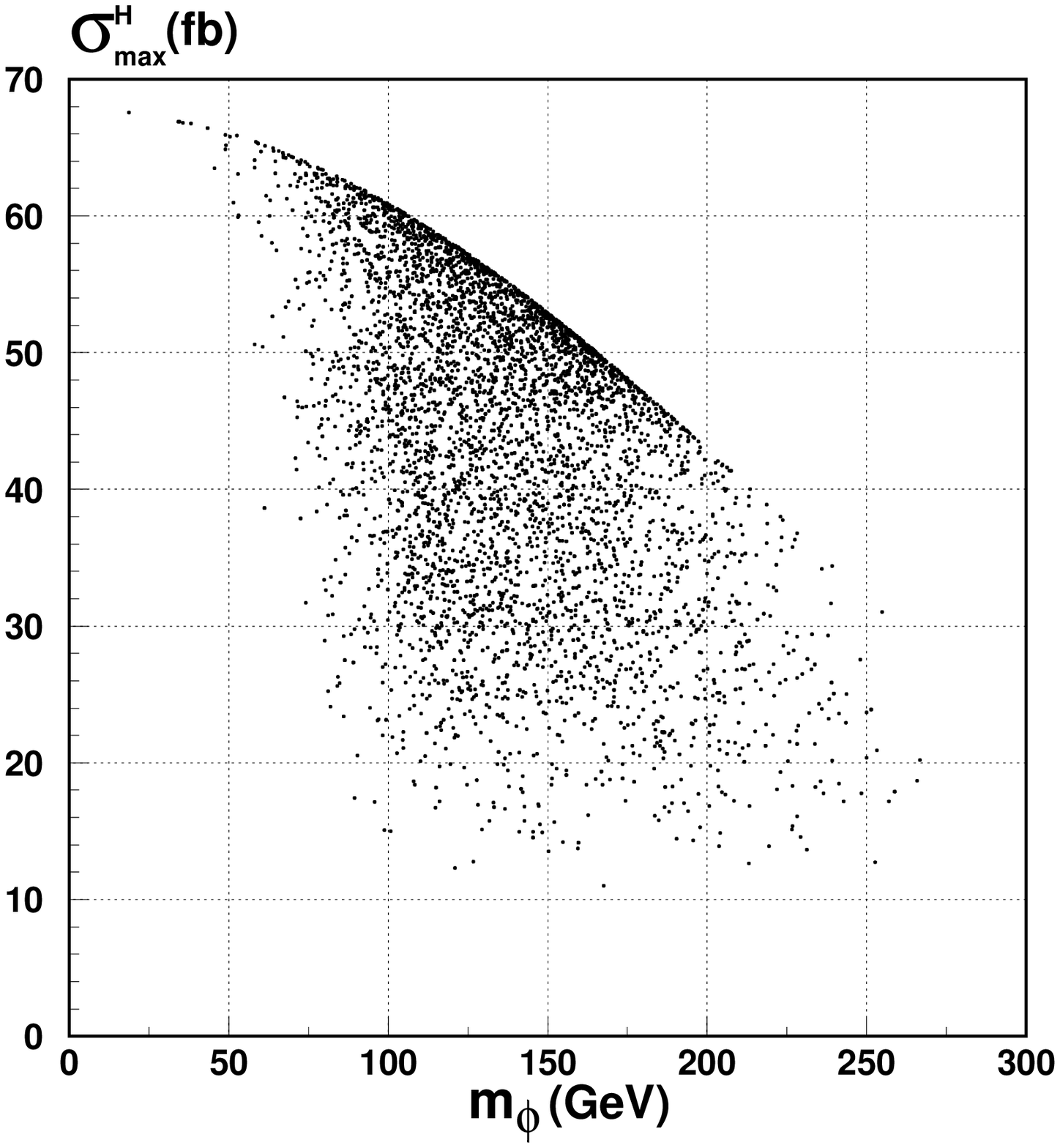}
\caption[plot]{For the same 5000 sets of parameter values as in Fig. 1a,
the larger one between the cross section for $\phi$ production and the cross section
for $S$ production via Higgs-strahlung process in $e^+e^-$ collisions with $\sqrt{s} = 500$ GeV
is plotted against $m_{\phi}$.}
\end{figure}
\renewcommand\thefigure{Fig. 4a}
\begin{figure}[t]
\epsfxsize=12cm
\hspace*{2.cm}
\epsffile{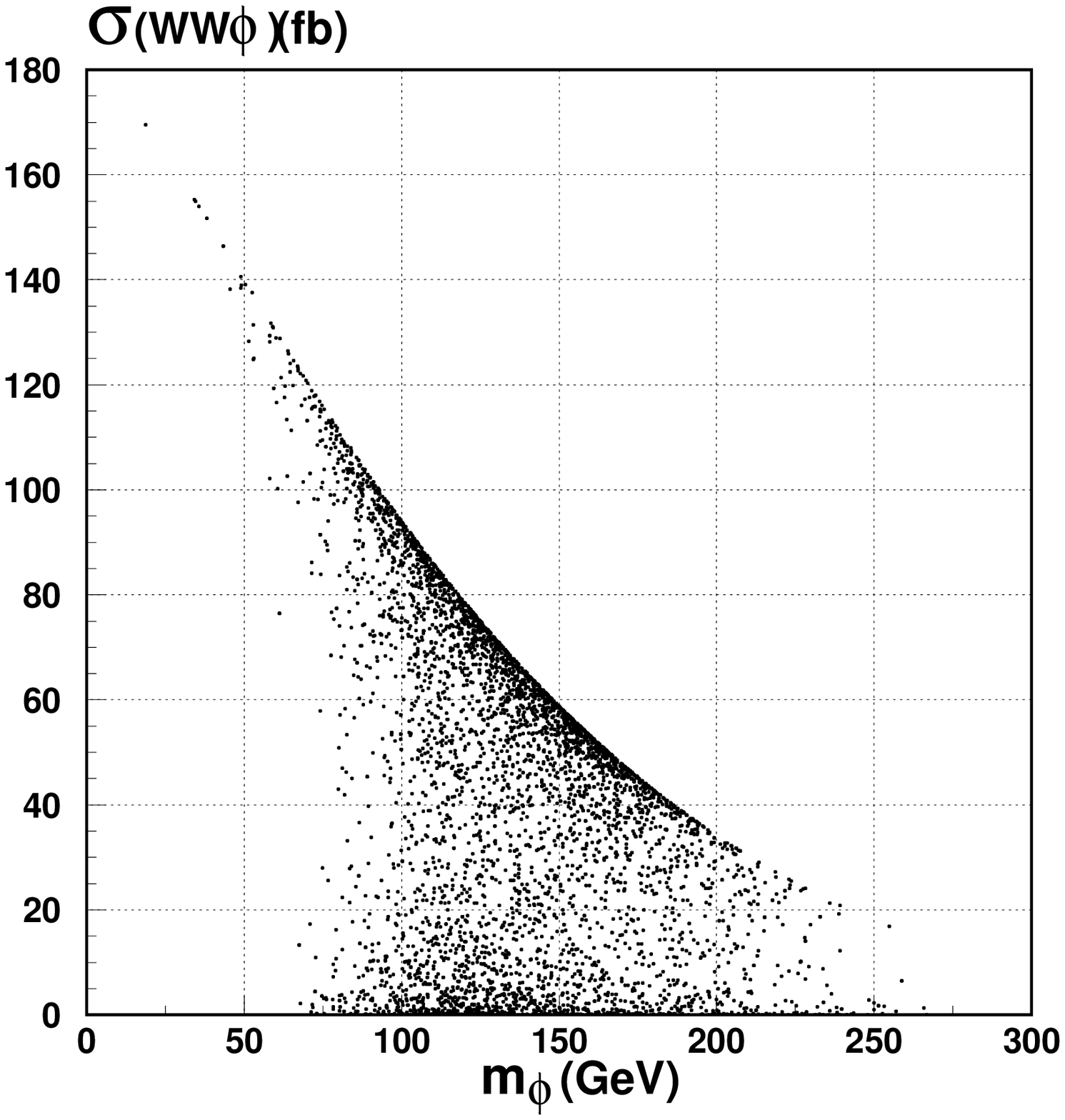}
\caption[plot]{The same as Figs. 3a, except for $WW$ fusion process instead
of the Higgs-strahlung process.}
\end{figure}
\renewcommand\thefigure{Fig. 4b}
\begin{figure}[t]
\epsfxsize=12cm
\hspace*{2.cm}
\epsffile{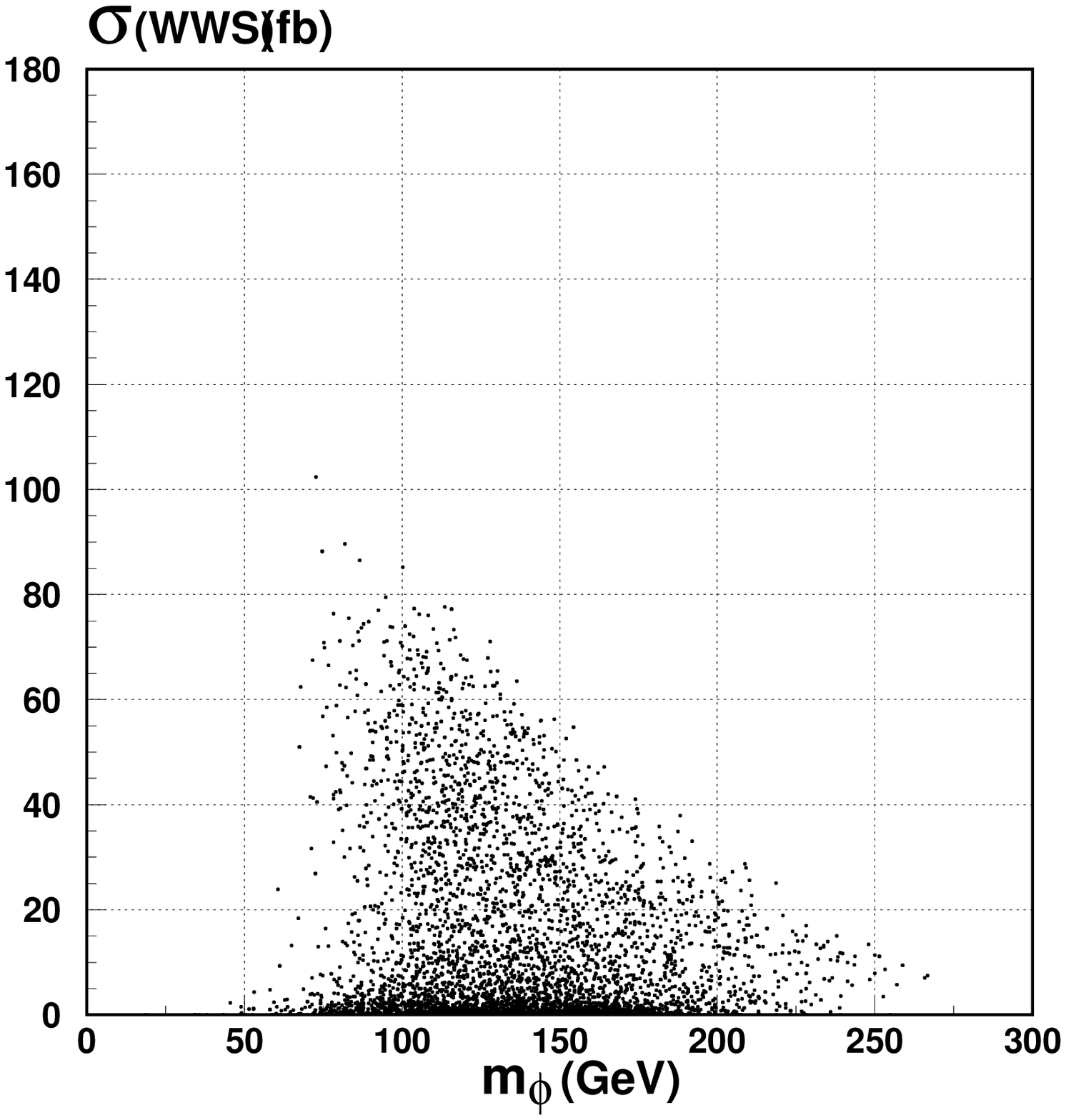}
\caption[plot]{The same as Figs. 3b except for $WW$ fusion process instead
of the Higgs-strahlung process.}
\end{figure}
\renewcommand\thefigure{Fig. 4c}
\begin{figure}[t]
\epsfxsize=12cm
\hspace*{2.cm}
\epsffile{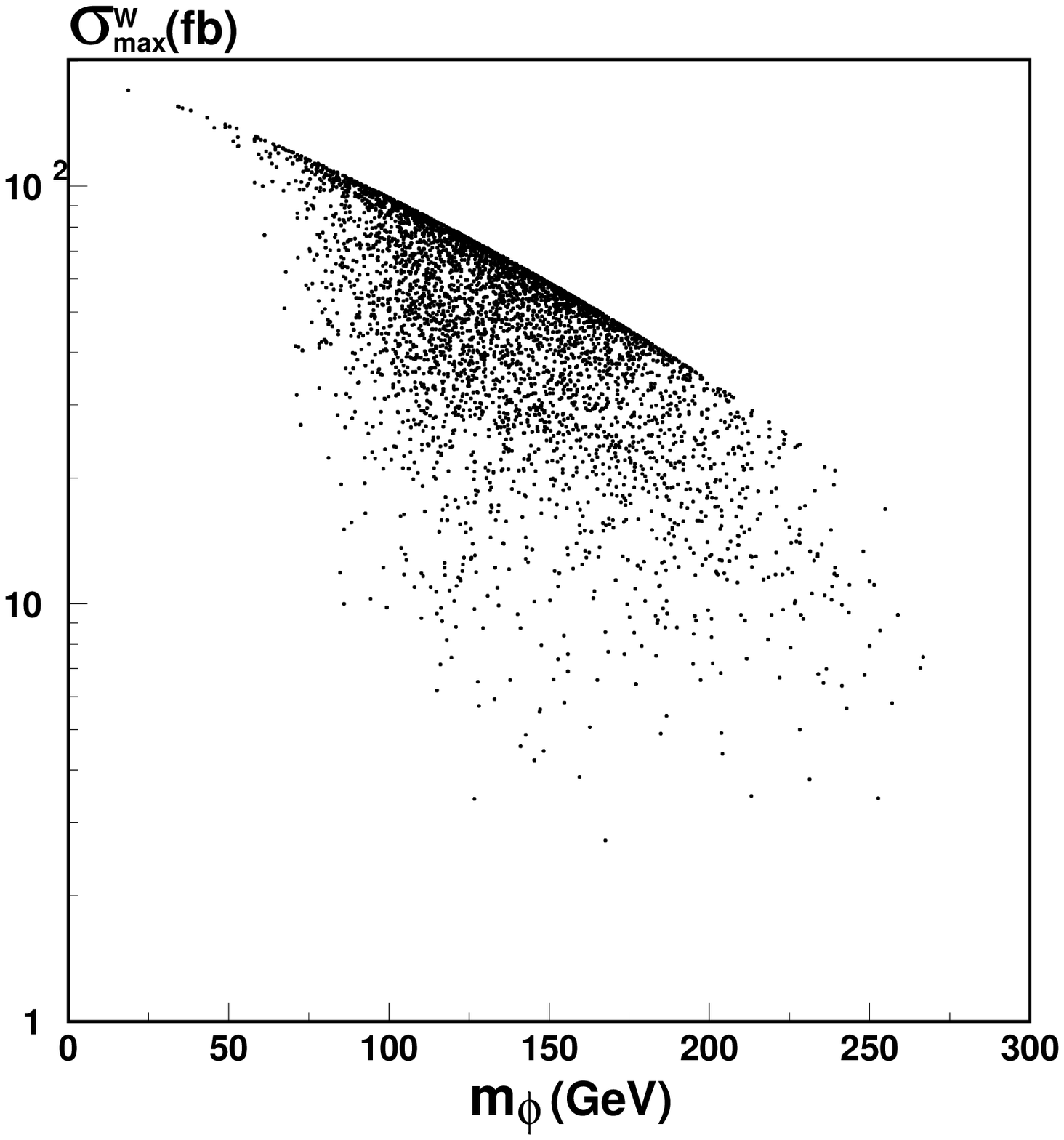}
\caption[plot]{The same as Figs. 3c except for $WW$ fusion process instead
of the Higgs-strahlung process.}
\end{figure}
\renewcommand\thefigure{Fig. 5a}
\begin{figure}[t]
\epsfxsize=12cm
\hspace*{2.cm}
\epsffile{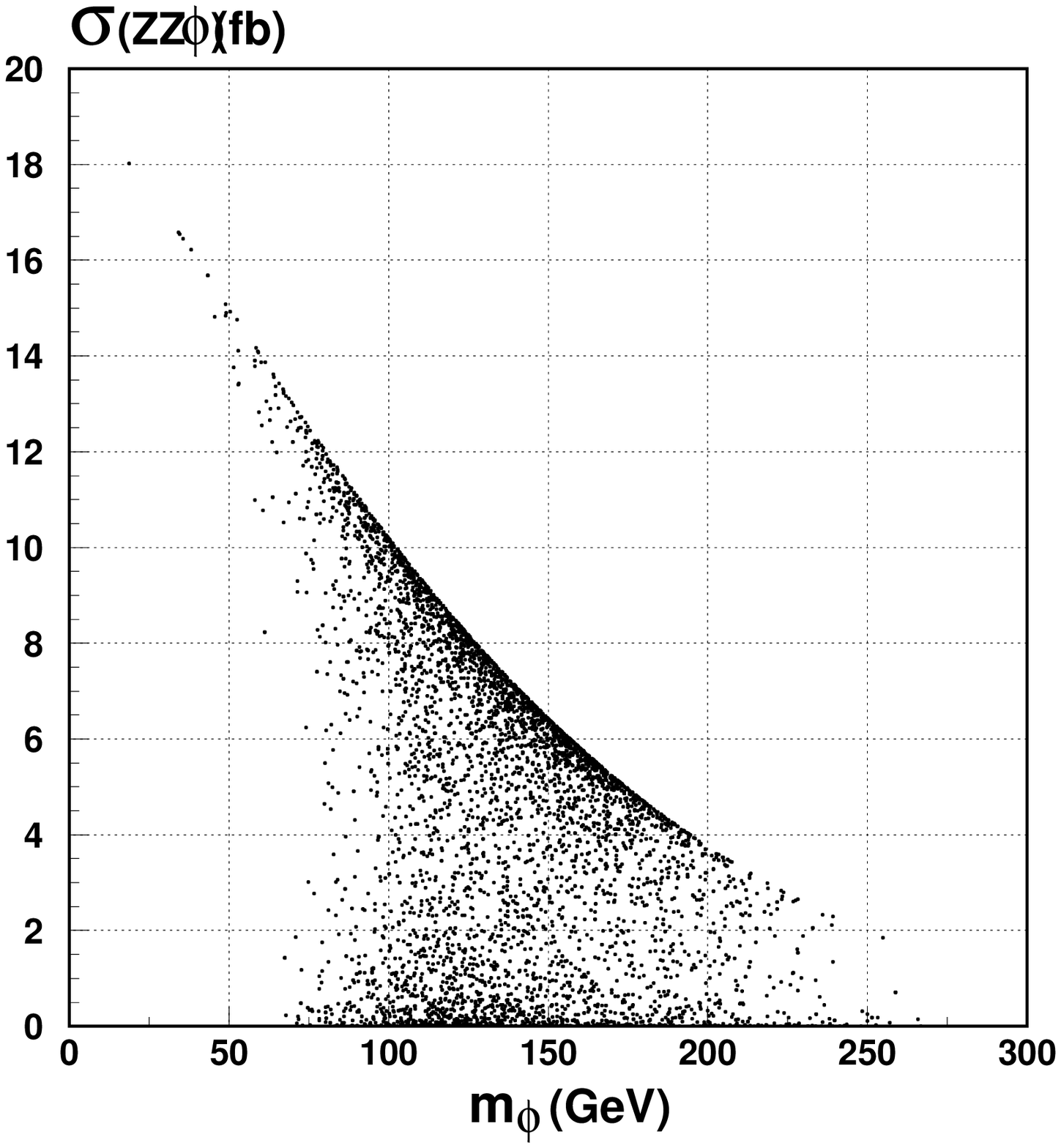}
\caption[plot]{The same as Figs. 3a, except for $ZZ$ fusion process instead
of the Higgs-strahlung process.}
\end{figure}
\renewcommand\thefigure{Fig. 5b}
\begin{figure}[t]
\epsfxsize=12cm
\hspace*{2.cm}
\epsffile{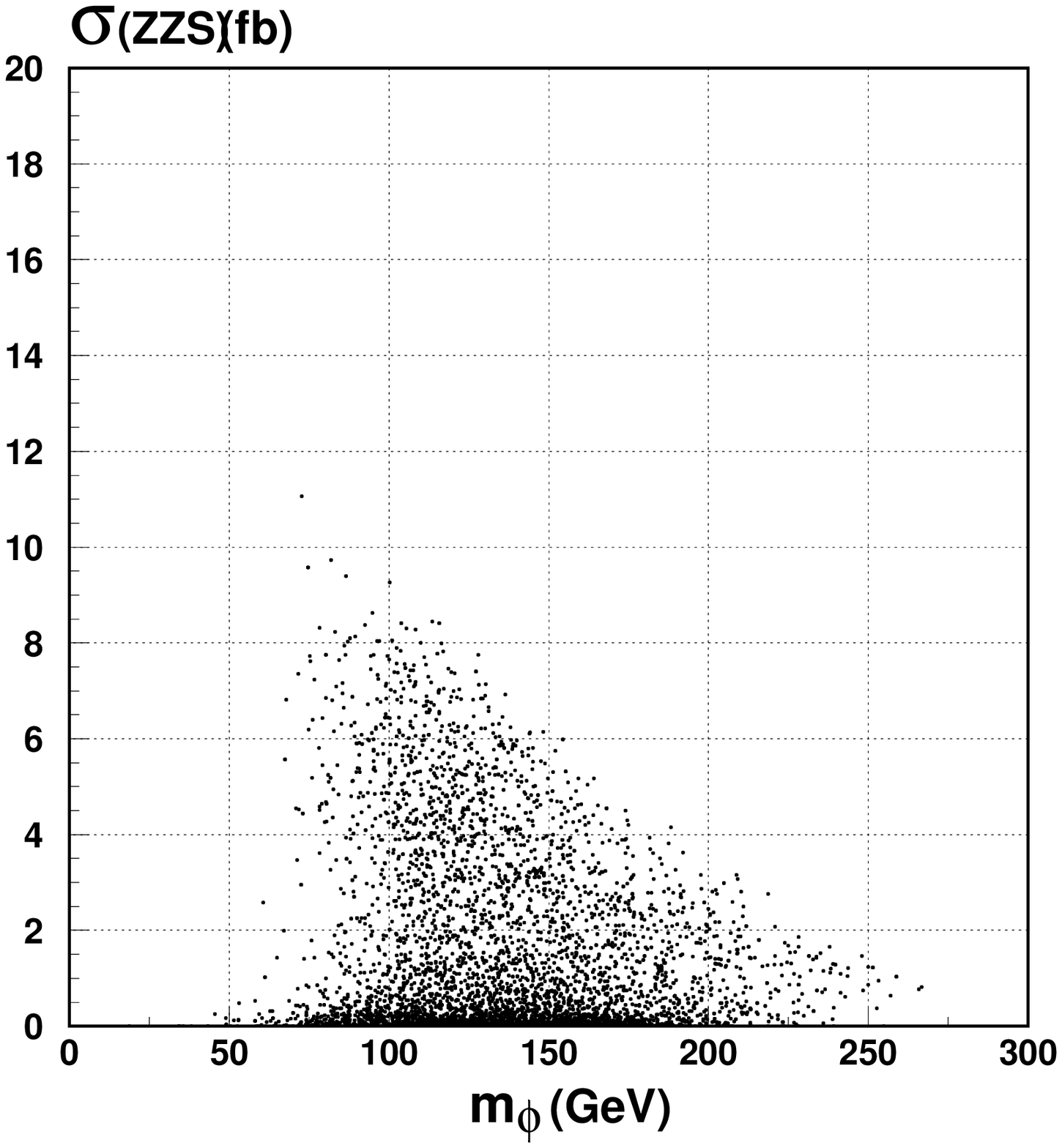}
\caption[plot]{The same as Figs. 3b, except for $ZZ$ fusion process instead
of the Higgs-strahlung process.}
\end{figure}
\renewcommand\thefigure{Fig. 5c}
\begin{figure}[t]
\epsfxsize=12cm
\hspace*{2.cm}
\epsffile{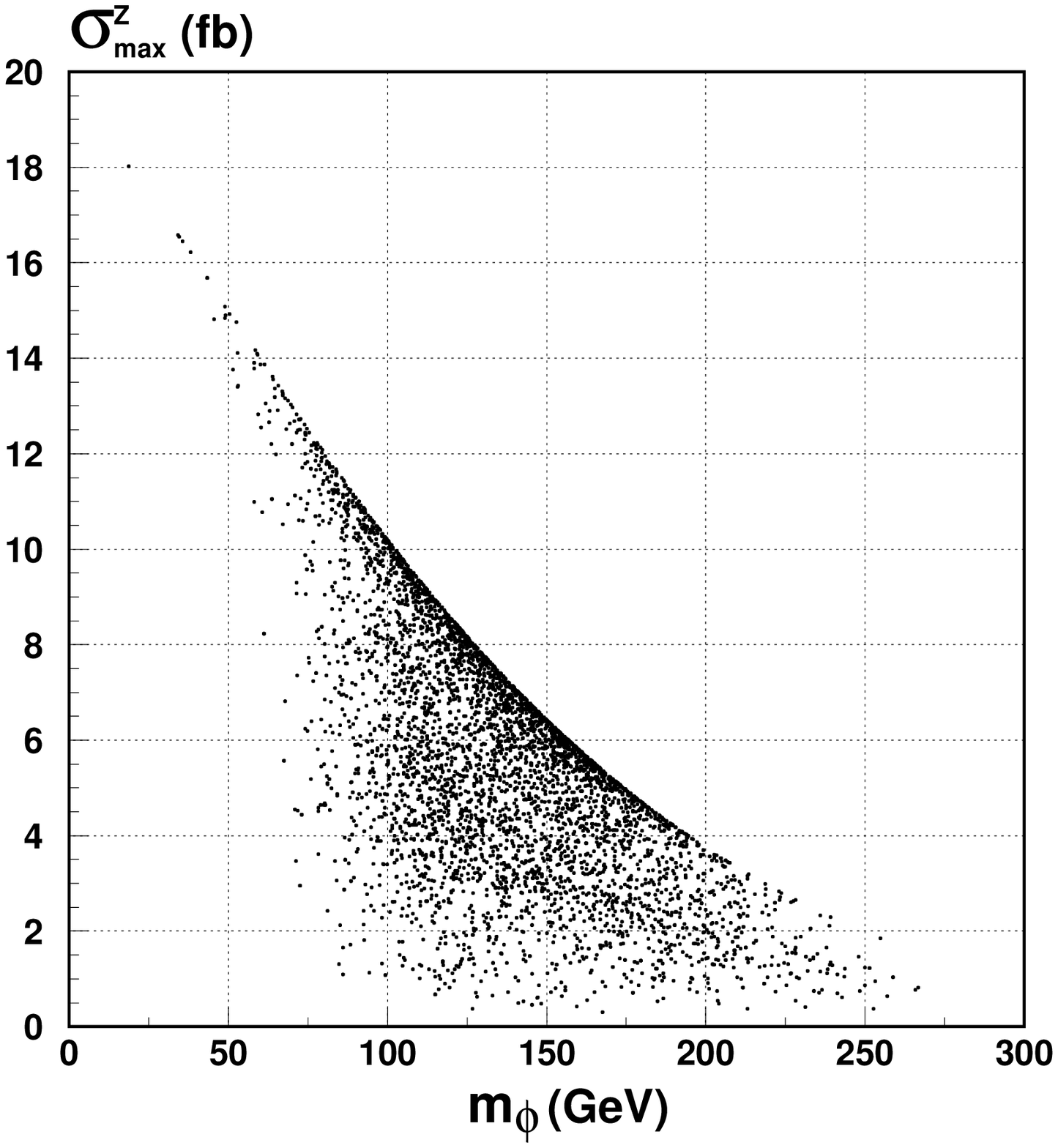}
\caption[plot]{The same as Figs. 3c, except for $ZZ$ fusion process instead
of the Higgs-strahlung process.}
\end{figure}


\begin{thebibliography}{99}

\bibitem{1}
V.A. Kuzmin, V.A. Rubakov, and M.E. Shaposhnikov, Phys. Lett. B {\bf 155}, 36 (1985);
M.E. Shaposhnikov, JETP Lett. {\bf 44}, 465 (1986);
    Nucl. Phys. B {\bf 287}, 757 (1987);
    Nucl. Phys. B {\bf 299}, 797 (1988);
L. McLerran, Phys. Rev. Lett. {\bf 62}, 1075 (1989);
N. Turok and J. Zadrozny, Phys. Rev. Lett. {\bf 65}, 2331 (1990);
    Nucl. Phys. B {\bf 358}, 471 (1991);
L.D. McLerran, M.E. Shaposhnikov, N. Turok, and M.B. Voloshin, Phys. Lett. B {\bf 256}, 451 (1991);
M. Dine, P. Huet, R.J. Singleton, and L. Susskind, Phys. Lett. B {\bf 257}, 351 (1991);
A.G. Cohen, D.B. Kaplan, and A.E. Nelson, Annu. Rev. Nucl. Part. Sci. {\bf 43}, 27 (1993);
M. Trodden, Rev. Mod. Phys. {\bf 71}, 1463 (1999);
A. Riotto and M. Trodden, Annu. Rev. Nucl. Part. Sci. {\bf 49}, 35 (1999).
\bibitem{2}
A.D. Sakharov, JETP Lett. {\bf 5}, 24 (1967).
\bibitem{3}
A.I. Bochkarev, S.V. Kuzmin, and M.E. Shaposhnikov, Mod. Phys. Lett. A {\bf 2}, 417 (1987);
    Phys. Lett. B {\bf 244}, 275 (1990);
M. Dine, R.G. Leigh, P. Huet, A. Linde, and D. Linde, Phys. Rev. D {\bf 46}, 550 (1992);
P. Arnold and O. Espinosa, Phys. Rev. D {\bf 47}, 3546 (1993);
Z. Fodor and A. Hebecker, Nucl. Phys. B {\bf 432}, 127 (1994);
K. Kajantie, M. Laine, K. Rummukainen, and M. Shaposhnikov, Phys. Rev. Lett. {\bf 77}, 2887 (1996);
F. Csikor, Z. Fodor, and J. Heitger, Phys. Rev. Lett. {\bf 82}, 21 (1999).
\bibitem{4}
A.I. Bochkarev, S.V. Kuzmin and M.E. Shaposhnikov, Phys. Rev. D {\bf 43}, 369 (1991);
G.W. Anderson and L.J. Hall, Phys. Rev. D {\bf 45}, 2685 (1992);
N. Turok and J. Zadrozny, Nucl. Phys. B {\bf 369}, 729 (1992);
G.F. Giudice, Phys. Rev. D {\bf 45}, 3177 (1992);
J.R. Espinosa, M. Quir\'os, and F. Zwirner, Phys. Lett. B {\bf 307}, 106 (1993);
K. Enqvist, K. Kainulainen, and I. Vilja, Nucl. Phys. B {\bf 403}, 749 (1993);
M. Pietroni, Nucl. Phys. B {\bf 402}, 27 (1993);
I. Vilja, Phys. Lett. B {\bf 324}, 197 (1994);
A.T. Davies, C.D. Froggatt, G. Jenkins, and R.G. Moorhouse, Phys. Lett. B {\bf 336}, 464 (1994);
A.T. Davies, C.D. Froggatt, and R.G. Moorhouse, Phys. Lett. B {\bf 372}, 88 (1996);
M. Carena, M. Quir\'os and C.E.M. Wagner, Phys. Lett. B {\bf 380}, 81 (1996);
D. Delepine, J.M. Gerard, R.G. Felipe, and J. Weyers, Phys. Lett. B {\bf 386}, 183 (1996);
B. de Carlos and J.R. Espinosa, Nucl. Phys. B {\bf 503}, 24 (1997).
J.M. Cline and P.A. Lemieux, Phys. Rev. D {\bf 55}, 3873 (1997);
M. Carena, M. Quir\'os and C.E.M. Wagner, Nucl. Phys. B {\bf 524}, 3 (1998);
M. Laine and K. Rummukainen, Phys. Rev. Lett. {\bf 80}, 5259 (1998);
    Nucl. Phys. B {\bf 535}, 423 (1998);
J.M. Cline and G.D. Moore, Phys. Rev. Lett.  {\bf 81}, 3315 (1998);
K. Funakubo, A. Kakuto, S. Otsuki, and F. Toyoda, Prog. Theor. Phys. {\bf 99}, 1045 (1998);
S.J. Huber and M.G. Schmidt, Eur. Phys. J. C {\bf 10}, 473 (1999);
K. Funakubo, Prog. Theor. Phys. {\bf 101}, 415 (1999);
M. Losada, Nucl. Phys. B {\bf 537}, 3 (1999);
M. Bastero-Gil, C. Hugonie, S.F. King, D.P. Roy, and S. Vempati, Phys. Lett. B {\bf 489}, 359 (2000);
A. Menon, D.E. Morrissey, and C.E.M. Wagner, Phys. Rev. D {\bf 70}, 035005 (2004);
S.W. Ham, S.K. Oh, C.M. Kim, E.J. Yoo, and D. Son, Phys. Rev. D {\bf 70}, 075001 (2004);
J. Kang, P. Langacker, T. Li, and T. Liu,  Phys. Rev. Lett. {\bf 94}, 061801 (2005);
M. Carena, A. Megevand, M. Quir\'os, and C.E.M. Wagner, hep-ph/0410352.
\bibitem{5}
J. Choi and R.R. Volkas, Phys. Lett. B {\bf 317}, 385 (1993);
J. Choi, Phys. Lett. B {\bf 345}, 253 (1995).
\bibitem{6} S. Coleman and E. Weinberg, Phys. Rev. D {\bf 7}, 1888 (1973).
\bibitem{7} L. Dolan and R. Jackiw, Phys. Rev. D {\bf 9}, 3320 (1974).
\bibitem{8} V. Barger, C.W. Chiang, J. Jiang, and T. Li, Nucl. Phys. B 705, 71 (2005).
\bibitem{9} LEP Higgs Working Group for Higgs boson searches, hep-ex/0107034.
\bibitem{10} X. Zhang, S.K. Lee, K. Whisnant, and B.L. Yong, Phys. Rev. D 50, 7042 (1993).
\bibitem{11} G.C. Branco, D. Delepine, D. Emmanuel-Costa, R. Gonzalez Felipe,
Phys. Lett. B 442, 229 (1998).

\end{thebibliography}
\end{document}